\documentclass[a4paper,preprint,superscriptaddress]{revtex4}
\usepackage{times}
\usepackage{amsmath}
\usepackage{graphicx}
\usepackage{amssymb}

\begin{document}

\title{Neutral and ionic dopants in helium clusters: interaction forces
for the $\mathrm{Li_{2}(a^{3}\Sigma_{u}^{+})-He}$ and $\mathrm{Li_{2}^{+}(X^{2}\Sigma_{g}^{+})-He}$
complexes.}

\author{E. Bodo}
\affiliation{Dept. of Chemistry and INFM, University of Rome {}``La Sapienza'',
Italy}

\author{F. A. Gianturco}
\thanks{Corresponding author: Dep. of Chemistry, University of Rome {}``La
Sapienza'', P. A. Moro 5, 00185, Rome, Italy. Fax: +39-06-49913305. }
\email{fa.gianturco@caspur.it}
\affiliation{Dept. of Chemistry and INFM, University of Rome {}``La Sapienza'',
Italy}

\author{E. Yurtsever}
\affiliation{Dept. of Chemistry, Koç University, Istanbul, Turkey}

\author{M. Yurtsever}
\affiliation{Dept. of Chemistry, Technical University of Istanbul, Turkey }

\begin{abstract}
The potential energy surface (PES) describing the interactions between
$\mathrm{Li_{2}(^{1}\Sigma_{u}^{+})}$ and $\mathrm{^{4}He}$ and
an extensive study of the energies and structures of a set of small
clusters, $\mathrm{Li_{2}(He)_{n}}$, have been presented by us in
a previous series of publications \cite{zb:bodo04,zb:bodo04-b,zb:bodo05-a}.
In the present work we want to extend the same analysis to the case
of the excited $\mathrm{Li_{2}}(a^{3}\Sigma_{u}^{+})$ and of the
ionized Li$_{2}^{+}(X^{2}\Sigma_{g}^{+})$ moiety. We thus show here
calculated interaction potentials for the two title systems and the
corresponding fitting of the computed points. For both surfaces the
MP4 method with cc-pV5Z basis sets has been used to generate an extensive
range of radial/angular coordinates of the two dimensional PES's which
describe rigid rotor molecular dopants interacting with one He partner. 
\end{abstract}
\maketitle

\section{Introduction}

The accurate evaluation of the interaction potential between a dopant
molecule or ion and a single $\mathrm{He}$ atom is an important step
in the study of small-size doped helium clusters because it can be
used to set up the total potential acting within them, an essential
first step in determining their structure and dynamics. Nano-sized
helium droplets in which a molecular or an atomic impurity has been
added by means of picking up techniques provide, indeed, a unique
environment for high precision spectroscopic studies of the molecular
solute \cite{cl:toennies04} especially for weakly bound molecular
species \cite{cl:stienkenmeier96,cl:higgins98,cl:stienkemeier01}
that are otherwise difficult to analyze. Among the many dopants that
have been experimentally and theoretically analyzed, the alkali metal
atoms and dimers have many interesting properties: from experimental
\cite{cl:stienkenmeier96,cl:stienkenmeier96-b,cl:stienkemeier01}
and theoretical \cite{cl:dalfovo94,zb:bodo04,zb:bodo05-a} evidence
it turns out that they normally reside on the surface of an $^{4}\mathrm{He}$
droplet, usually forming a slight {}``dimple'' on that surface.
When two or more alkali atoms are attached to the droplet they eventually
meet on its surface forming a singlet or a triplet molecule. Spectroscopic
identification of the molecules shows \cite{cl:higgins98} that dimers
in the triplet state outnumber the singlet molecules by a factor up
to 10$^{5}$. This is probably due to the differences in stability
between the triplet dimers and the singlet molecules \cite{cl:higgins98}
whereby the energy release due to the formation of a singlet molecule
is more likely to lead to the detachment of the molecule and to the
almost complete disappearance of this kind of dopant. A similar behaviour,
i.e. one in which high spin states are preferred, has been noticed
also in the formation of small clusters of alkali atoms grown on helium
droplet surfaces \cite{cl:schulz04}: the very small binding energies
to the surface allows for a spontaneous {}``selection'' of the cluster
in higher spin states (it is worth pointing out that, in these experiments,
lithium atoms were shown to form only dimers while heavier alkalis
produced also larger structures). Very recently experiments have been
also performed on alkali atoms in $^{3}\mathrm{He}$ droplets \cite{cl:mayol04}
where both theoretical predictions and experiments \cite{cl:stienkenmeier04}
have shown a very similar behavior to that of the bosonic moieties.
The possible formation of cold heteronuclear molecules has attracted
the attention of researchers in the last few years (see for example
the special volume edition in which Ref. \cite{cl:mudrich04} can
be found) because of the possibility of studying the effects of polar
interactions in cold or even quantum degenerate molecular samples:
helium droplets may therefore allow the spectroscopic study of weakly
bound (triplet) heteronuclear dimers. 

In all the experiments mentioned above, the helium droplet is used
as a convenient, but inert matrix in which weakly bound molecular
species are accumulated and studied. Helium droplets, however, can
also be used as microscopic cryo-reactors which actively initiate
series of chemical reactions. Very recently, in fact, it has been
shown possible to follow chemical reactions inside a helium droplet
\cite{cl:farnick05} using, among others, simple diatomic dopants.
The chain of reactions is usually initiated by the ionization of one
of the helium atoms \cite{cl:farnick05} with an ensuing charge transfer
\cite{cl:xie03,zb:bodo04-i,cl:scifoni04} process that moves the charge
onto the impurity because its ionization potential is usually lower
than He. The impurity gives rise, in turn, to a variety of further,
secondary chemical reactions: the excess energy due to any exothermic
reaction is rapidly thermalized within the droplet itself by helium
atom evaporation. The rapid quenching of the internal degrees of freedom
of the partner species afforded by the superfluid environment of the
droplets has a profound effect on the various branching reactions
and, in some cases, can ultimately stabilize different intermediate
complexes which would instead only remain as transient species in
the gas phase. The latter condition therefore becomes an interesting
feature of such nanoscopic cryostats in the sense that any further
step initiated by ionization of the cluster can follow specific pathways
which are likely to be different from those under the corresponding
gas phase conditions. 

Concerning the experiments involving neutral dopants, we believe that
the microscopic structure of the helium surface-dimer complex is a
key piece of information for the full understanding of the experimental
data. In this work we continue the analysis of the lithium-He system
by presenting an accurate Potential Energy Surface (PES) for the dimer
in the triplet state $\mathrm{Li_{2}}(^{3}\Sigma_{u}^{+})$ and one
Helium atom. 

When moving to the problem of ionic dopants, and therefore to the
fascinating possibility of conducting ionic reactions at low temperatures
in the droplet, the study of the microscopic structure of the solvation
site around the ionic molecule may also help the understanding of
the possible reactive pathways which are eventually followed. This
is the reason why, together with the previously mentioned PES, we
also want to present here similar calculations on the $\mathrm{Li_{2}^{+}(^{2}\Sigma_{g}^{+})-He}$
system. These two PES's may be considered as the building blocks of
the complex interactions at play in the He droplets or in smaller
helium clusters: they describe two rather extreme situations of either
a weak, dispersion-type interaction or of a much stronger, markedly
orientational ionic PES. Although the forces at play in the large
systems may be extremely difficult to compute, a simplified approach
where the total potential is obtained as a sum of two-body potentials
represents a realistic route because the three-body forces and the
higher order terms, especially for the neutral system, provide fairly
negligible contributions to the total energies. For their possible
relevance in ionic systems see e.g. Refs. \cite{zzb:bodo05-accepted1}
and \cite{cl:lenzer01} for further details.

\section{The ab-initio calculations}

Both surfaces have been calculated using the MP4(SDTQ) method (without
freezing the lithium core) with the cc-pV5Z basis set employed within
Gaussian03 \cite{prg:gaussian98}. All energies have been corrected
for BSSE using the Counterpoise procedure \cite{ab:boys70}. Since,
in either case, the distortion due to the helium atom over the electronic
structure of the molecule is rather weak, we have decided to keep
the molecule at its equilibrium geometry. The optimal geometry of
the molecular species was determined by a separate MP4 optimization
on both the triplet Li$_{2}$ and ionic $\mathrm{Li}{}_{2}^{+}$.
The internuclear distances that we have used are $r=4.175$~\AA~
and $r=3.11$ ~\AA~ for Li$_{2}$ and Li$_{2}^{+}$ respectively.
For both surfaces we have used a grid of Jacobi coordinates with different
radial geometries and 19 angles from 0° to 90°. For the Li$_{2}^{+}-$He
case we have calculated for each angle a number of points between
50 and 64 ($2.5\le R\le12.0$~\AA~) for a total number of 1,135.
For the Li$_{2}-$He system the number of radial points for each angle
varied from 30 to 80 ($2.0\le R\le14.0$~\AA~) for a total of 847
points.

\section{The Fitting Procedure}

In order to make it easier to employ the two potential energy surfaces
in further studies,we have decided to obtain a suitable analytical
fitting of the raw points using a non-linear fitting procedure based
on the minimization of the square deviation, as obtained by means
of the efficient Levenberg-Marquadt method \cite{bk:numrec-book}.

The full interaction can be written as
\begin{equation}
V_{tot}=V(R_{a},\theta_{a})+V(R_{b},\theta_{b})+V_{LR}(R,\theta)
\end{equation}
where the coordinates are those reported for clarity in
Figure 1.
\begin{figure}
\begin{center}\includegraphics[width=0.50\textwidth]{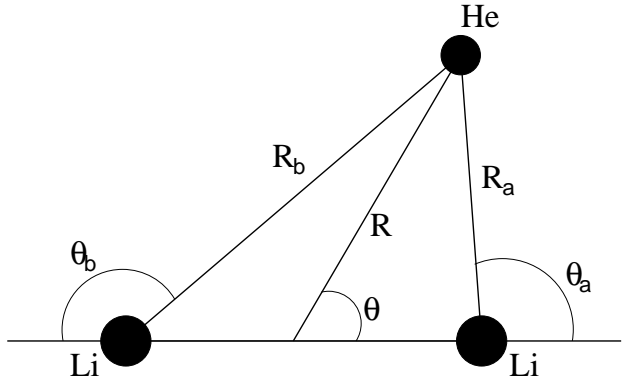}\end{center}
\caption{Coordinates used in the fitting formula}
\end{figure}
The first two contributions represent the anisotropic interactions
at short range and are written in terms of Legendre polynomials
\begin{equation}
V(R_{a},\theta_{a})=\sum_{n=0}^{nmax}\sum_{l=0}^{lmax}R_{a}^{n}\exp(-\beta R_{a})
\cdot P_{l}(cos\vartheta_{a})\cdot C_{nl}^{a}
\end{equation}
where, given the symmetry of the homonuclear molecule $C_{nl}^{a}=C_{nl}^{b}$.
The long-range contribution is instead expressed in Jacobi
coordinates and given by
\begin{equation}
V_{LR}(R,\theta)=\sum_{N}\sum_{L=0}^{N-4}\frac{f_{N}(\beta R)}
{R^{N}}\cdot P_{L}(cos\vartheta)\cdot C_{NL}^{LR}
\end{equation}
where the $f_{N}$ damping functions are those defined within
the well-known Tang-Toennies empirical potential modeling \cite{cl:kleinekatofer-96}.%
\begin{figure}
\includegraphics[width=1.0\textwidth]{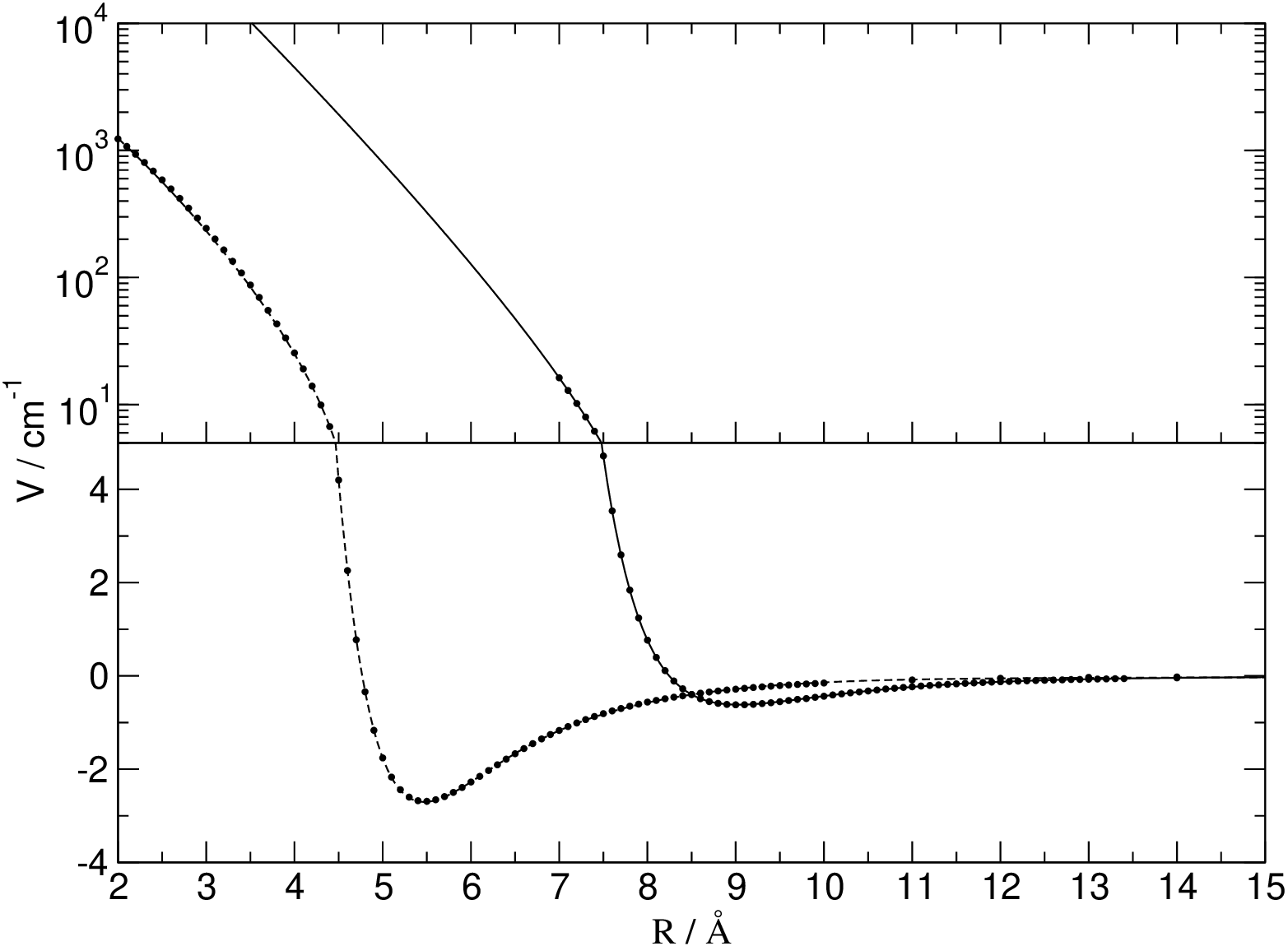}
\caption{Potential energy curves for Li$_{2}(^{3}\Sigma)$-He at $\theta=0°$
(solid line) and $\theta=90°$(dashed line) as obtained from our fitting.
Filled-in circles are ab-initio energies. }
\end{figure}
\begin{figure}
\includegraphics[width=1.0\textwidth]{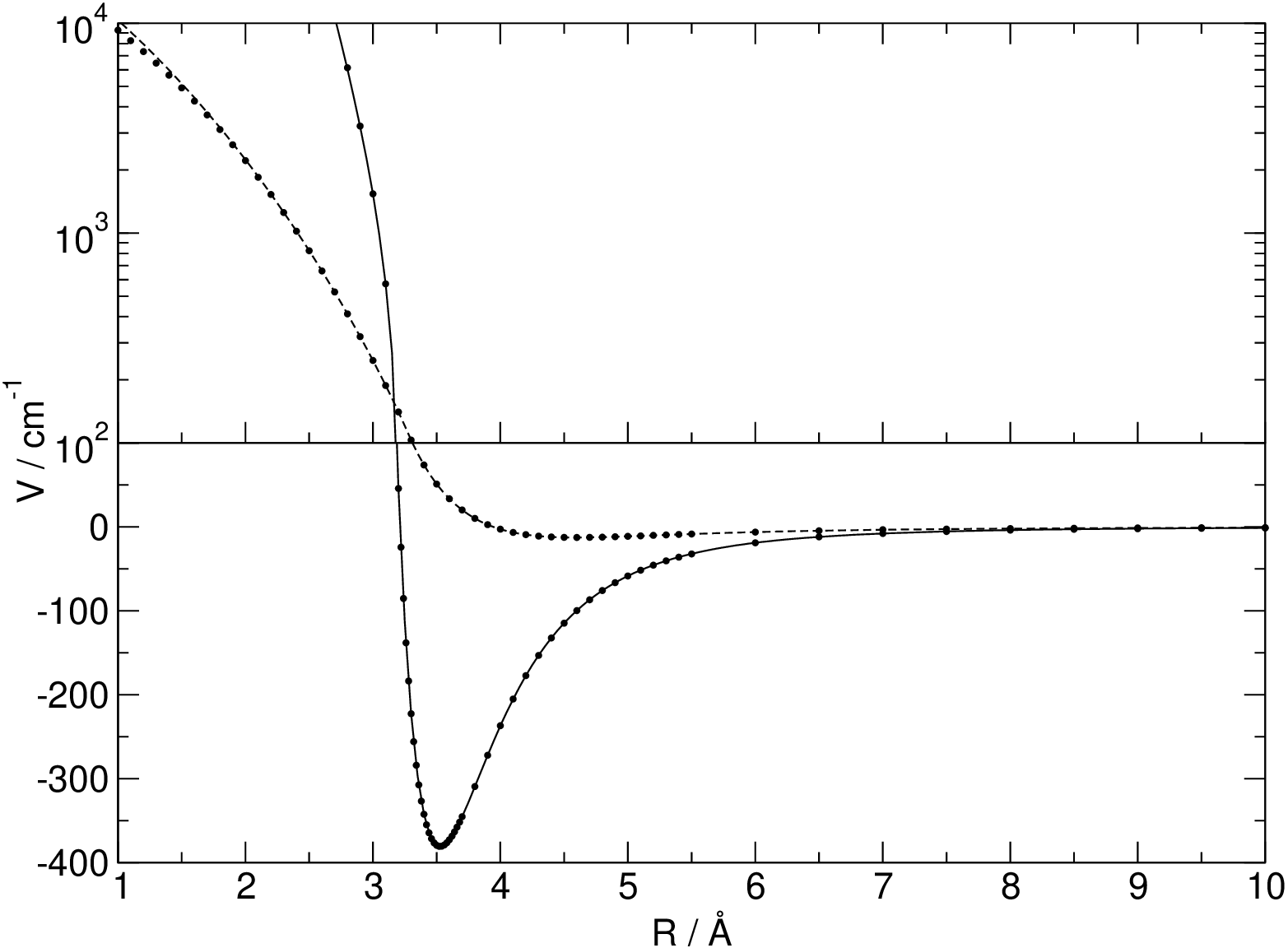}
\caption{Potential energy curves for Li$_{2}^{+}$-He at $\theta=0°$ (solid
line) and $\theta=90°$(dashed line) as obtained from our fitting.
Filled-in circles are ab-initio energies. }
\end{figure}

For the neutral system (Li$_{2}$-He) we have employed $nmax=4$ and
$lmax=6$ and we have limited the long range expression to the value
$N=6$, i.e. the latter was simply given by two terms
 \[ V_{LR}=f_{6}(\beta R)\cdot R^{-6}\cdot[P_{0}(cos\theta)C_{60}^{LR}+P_{2}(cos\theta)C_{62}^{LR}].\]
 This very simple analytical representation of the entire set of unweighted
847 ab-initio points yielded a standard deviation of 0.0046 cm$^{-1}$.
For the ionic system (Li$_{2}^{+}$-He) we have employed a more flexible
representation given by $nmax=8$ and $lmax=4$ and two long range
anisotropic terms: $C_{4}$ and $C_{6}$. The latter terms were employed
in order to represent correctly the long range multipolar expansion
of a neutral, polarizable atom interacting with a point-like charge
ion. The long range expansion therefore becomes:
\[V_{LR}=f_{4}(\beta R)\cdot R^{-4}\cdot
[P_{0}(cos\theta)C_{40}^{LR}]+f_{6}(\beta R)
\cdot R^{-6}\cdot[P_{0}(cos\theta)C_{60}^{LR}+P_{2}(cos\theta)C_{62}^{LR}].\]
The total number of 1,135 points was reduced to 1,065 by excluding
the repulsive energies with values above 5,000 cm$^{-1}$. A simple
weight function $W(E)$ was used in order to obtain a better representation
of the interaction region and to reduce the importance of the repulsive
geometries. Hence we drop the weight of the points in the less important
highly repulsive regions of the interaction
\[ W(E)=\begin{cases}
1 & \textrm{for }E\le1,000\\
|E|^{-1} & \textrm{for }1,000\le E\le5,000\end{cases}\]
The final standard deviation was 0.073 cm$^{-1}$. As an example of
the accuracy of the two fittings we report in Figure 2 (neutral system)
and Figure 3 (ionic system) two selected cuts through the PES's at
$\theta$=0° and $\theta$=90° and compare them with the ab-initio
points. As is evident from the figures, we are able to describe with
good accuracy both the attractive region and the lowest section of
the repulsive wall. In either case, however, we were not able to provide
reliable fitting values for $R\le1.0$~\AA~. All the coefficients
are reported in Table I and Table II. Fortran77 subroutines are available
on request from the authors. 

\begin{table}

\caption{Coefficients for the Fitting of Li$_{2}(^{3}\Sigma)$-He}

\begin{center}\begin{tabular}{|c|c|c|c|c|c|c|c|}
\hline 
n l&
$C_{nl}$&
n l&
$C_{nl}$&
n l&
$C_{nl}$&
n l&
$C_{nl}$\\
\hline
\hline 
0 0&
22763.19&
1 3&
-122598.06&
2 6&
-825.49&
4 2&
1291.90\\
\hline 
0 1&
167466.30&
1 4&
57319.62&
3 0&
-2718.38&
4 3&
-513.93\\
\hline 
0 2&
-171109.44&
1 5&
-11157.70&
3 1&
1618.93&
4 4&
69.19\\
\hline 
0 3&
116907.07&
1 6&
567.34&
3 2&
672.89&
4 5&
16.31\\
\hline 
0 4&
-44321.81&
2 0&
-5532.20&
3 3&
-2363.22&
4 6&
-7.15\\
\hline 
0 5&
-675.99&
2 1&
31713.40&
3 4&
2099.82&
N L&
$C_{NL}^{LR}$\\
\hline 
0 6&
3002.69&
2 2&
-43241.69&
3 5&
-816.34&
6 0&
-560011.07\\
\hline 
1 0&
5967.45&
2 3&
37506.82&
3 6&
157.05&
6 2&
105854.86\\
\hline 
1 1&
-143899.90&
2 4&
-20326.82&
4 0&
648.35&
-&
$\beta$\\
\hline 
1 2&
162737.35&
2 5&
5713.51&
4 1&
-1524.99&
-&
1.09998326\\
\hline
\end{tabular}\end{center}
\end{table}

\begin{table}

\caption{Coefficients for the Fitting of Li$_{2}(^{3}\Sigma)$-He}

\begin{center}\begin{tabular}{|c|c|c|c|c|c|c|c|}
\hline 
n l&
$C_{nl}$&
n l&
$C_{nl}$&
n l&
$C_{nl}$&
n l&
$C_{nl}$\\
\hline
\hline 
0 0&
1794639.56&
2 3&
-26905991.00&
5 1&
7206051.75&
7 4&
-4773.84\\
\hline 
0 1&
-4891313.47&
2 4&
-5708309.41&
5 2&
-7598549.02&
8 0&
4440.63\\
\hline 
0 2&
9692348.34&
3 0&
-7120382.44&
5 3&
4029614.92&
8 1&
-12773.88\\
\hline 
0 3&
-6003108.42&
3 1&
33556992.50&
5 4&
440943.65&
8 2&
12647.36\\
\hline 
0 4&
-991282.46&
3 2&
-41302351.90&
6 0&
529353.73&
8 3&
-7220.96\\
\hline 
1 0&
-3645612.91&
3 3&
21971925.50&
6 1&
-1650913.80&
8 4&
760.89\\
\hline 
1 1&
20414626.80&
3 4&
4504104.73&
6 2&
1675760.54&
N L&
$C_{NL}^{LR}$\\
\hline 
1 2&
-33068152.00&
4 0&
5071552.76&
6 3&
-909621.15&
4 0&
-10627.48\\
\hline 
1 3&
19171419.60&
4 1&
-19611375.40&
6 4&
-32113.64&
6 0&
+43087.50\\
\hline 
1 4&
3756985.86&
4 2&
22000474.60&
7 0&
-73683.34&
6 2&
-212414.50\\
\hline 
2 0&
6019745.96&
4 3&
-11557831.20&
7 1&
217607.26&
-&
$\beta$\\
\hline 
2 1&
-35226804.20&
4 4&
-1959452.27&
7 2&
-216505.97&
-&
3.01084652\\
\hline 
2 2&
48899465.50&
5 0&
-2124762.77&
7 3&
120820.53&
-&
-\\
\hline
\end{tabular}\end{center}
\end{table}

\section{Results and discussion}

\subsection{The Potential energy surfaces}

A 3D representation of the two PES's is presented in Figure 4 as isoenergetic
contour plots. 
\begin{figure}
\includegraphics[width=0.75\textheight]{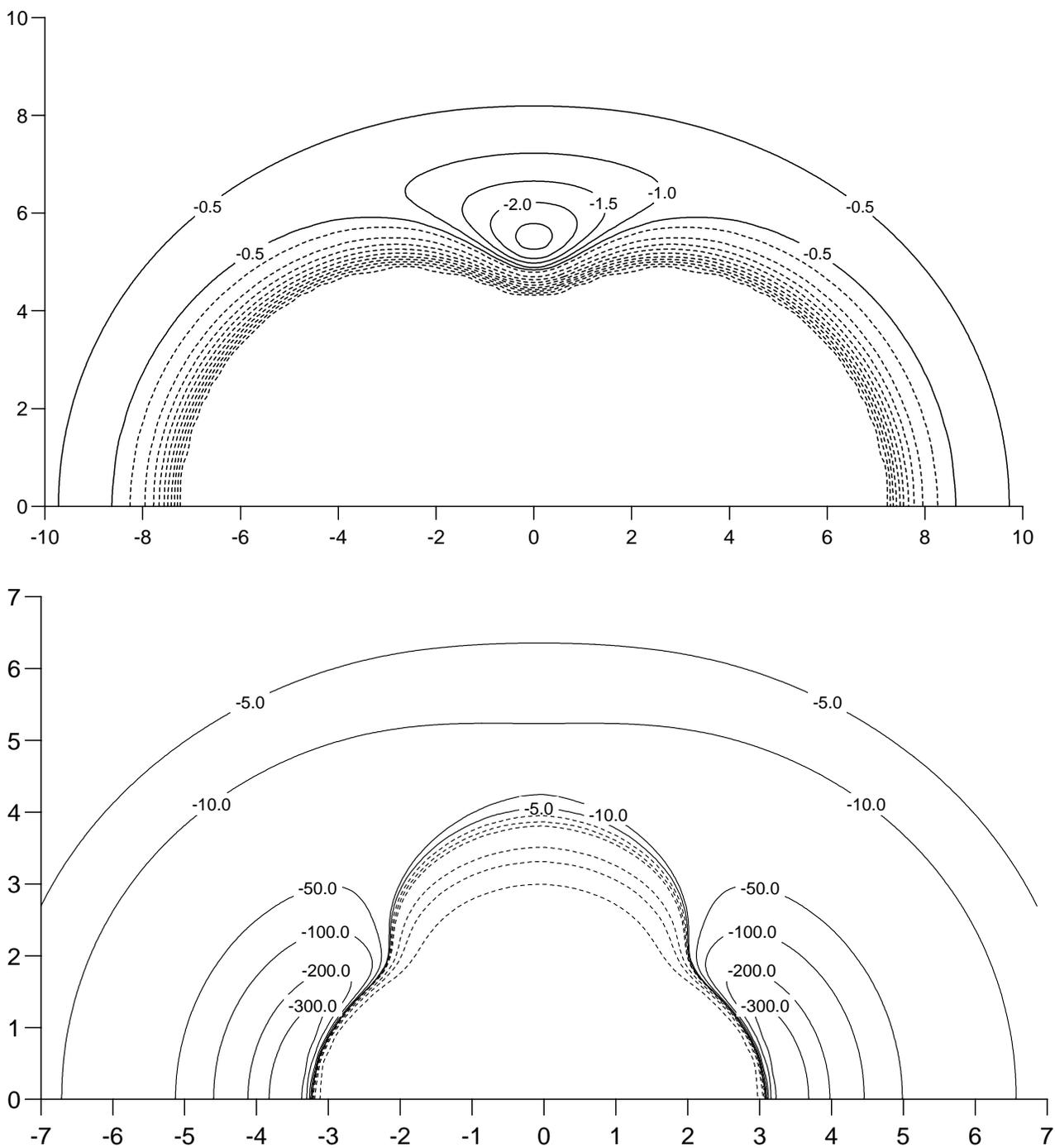}
\caption{Contour plots of the PES's for $\mathrm{Li_{2}}(^{3}\Sigma_{u}^{+})$-He
(upper panel) and $\mathrm{Li_{2}^{+}(^{2}\Sigma_{g}^{+})-He}$ (lower
panel). Dashed contours are positive energy isolines.}
\end{figure}
As can be seen from the figure, the two PES's are completely different
in shape and in strength, as one could easily have expected. The triplet
interaction is very weak (very similar to the one we have already
calculated for the singlet molecule in Ref. \cite{zb:bodo04,zb:bodo05-a})
and has its absolute minimum at $\theta$=90° with an interaction
energy of $\sim-2.4$ cm$^{-1}$. This is a very different anisotropy
with respect to that exhibited by the singlet dimer \cite{zb:bodo04,zb:bodo05-a},
where the collinear configuration provided the global PES minimum
energy and it also exhibits deeper well values than those of the former
case. The extreme weakness of the interaction of Li$_{2}$ with He
is indirectly supported by the experimental findings \cite{cl:stienkemeier01}
which indicate that triplet molecules also reside on the surface of
the droplet and are not efficiently solvated by the helium atoms. 

Given the weakness of the calculated interaction it would be also
be interesting to provide a modification of it that makes use of semi-empirical
guesses for the two-body contribution due to the Li-He interactions.
We can, in fact, modify the calculated interaction energies by employing
2-body semiempirical potentials along the lines which we have already
followed for the singlet molecule in Ref. \cite{zb:bodo05-a}. It
is, in fact, possible to produce rather simply a semi-empirical potential
$V'(R,\theta)$ by using the formula
\[V'(R,\theta)=V(R,\theta)-V^{ab}(R_{a})-V^{ab}(R_{b})+V^{sm}(R_{a})+V^{sm}(R_{b})\]
where $V^{sm}(R_{a,b})$ is the semi-empirical potential proposed
by Toennies and coworkers \cite{cl:kleinekatofer-96} for the dimer
Li-He, $V^{ab}(R_{a,b})$ is a simple fitting of the ab-initio points
calculated by us at the same level of accuracy (MP4/cc-pV5Z) for the
Li-He diatomic curve and $R_{a}$, $R_{b}$ are the two Li-He distances
within the trimer. Our ab-initio data, the fitting curve and the model
potential of Ref. \cite{cl:kleinekatofer-96} for the Li-He pair are
reported in Figure 5. The resulting full potential energy surface
$V'(R,\theta)$ calculated with the model potential is stronger than
the one from purely ab-initio data, although the increase in well
depth is still not sufficient to allow for possible solvation of the
molecular impurity inside the helium clusters (see also our similar
conclusions for the singlet Li$_{2}$ in Ref. \cite{zb:bodo05-a}).
\begin{figure}
\includegraphics[width=1.0\textwidth]{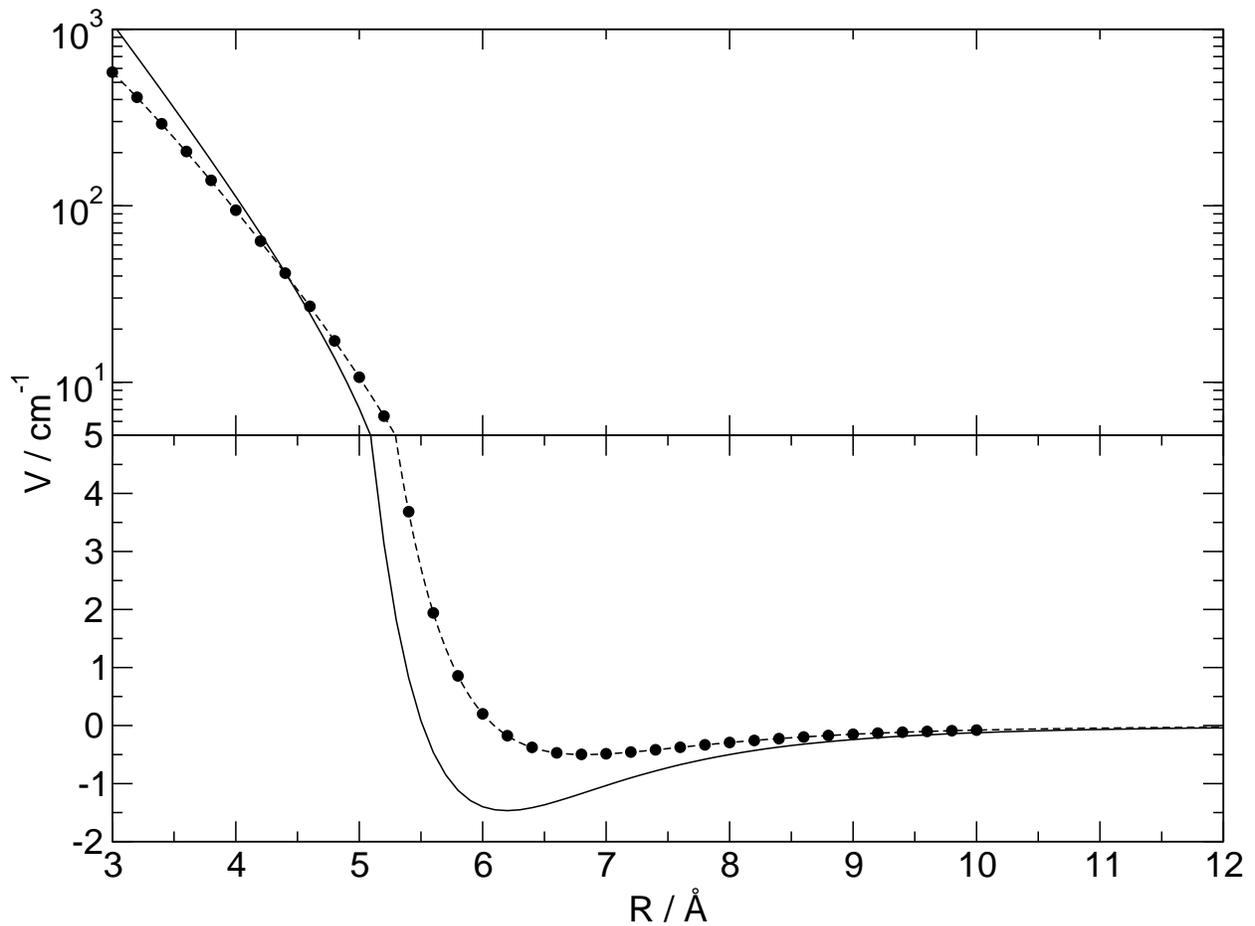}
\caption{Potential energy curves for Li-He pair. Filled in circles: present
ab initio data; solid line: model potential of Ref. \cite{cl:kleinekatofer-96};
dotted line: fitting of calculated ab-initio data using the following
formula:$V(R)=\sum_{n=0}^{2}a_{n}R^{n}e^{-bR}-f_{6}(bR)C_{6}R^{-6}$
with these parameters: $a_{0}$=271476, $a_{1}$=-33951.3, $a_{2}$=-1.11528$\cdot10^{6}$,
$b$=1.76229 and $C_{6}$=77803.7}
\end{figure}

The ionic interaction reported in Figures 3 and 4, on the other hand,
is much stronger and presents its minimum at $\theta$=0°. Here the
situation is completely different and, although no calculations have
been carried out as yet for the cluster structures, it is not difficult
to see that the Li$_{2}^{+}$ impurity would be strongly solvated
in liquid helium droplets and therefore it is likely to get localized
at the droplet center with a solvation shell strongly bound to it
(for analogous situations one should look at the existing results
on the ionic Li or Na impurities described in Refs. \cite{cl:galli02,zzb:bodo05-accepted1}).

\subsection{Estimating the 3-body effects in the ionic system}

As we have pointed out in the introduction, in order to treat small
and medium sized clusters with 2-50 helium atoms, one should be able
to rapidly evaluate the total potential energy acting within them.
One of the most used and successful approaches consists in approximating
the total interaction as a sum of 2-body terms, initially neglecting
any non-separable 3-body contribution. While this has been proved
to be a very accurate procedure for doped helium clusters with a neutral
impurity, it may represent a source of error in ionic clusters where
non-separable interactions among the induced multipoles in the first
solvation shell may be important. However, it has been recently shown
in various works on ionic dopants in rare-gas clusters including helium
that these effects are small and should not, when included, alter
substantially the geometries or the energies of the clusters (for
a general discussion with anionic dopants see refs. \cite{cl:lenzer01}
and \cite{cl:sebastianelli03,cl:sebastianelli03-b}, for positive
ions see Refs. \cite{zzb:bodo05-accepted1,zb:bodo05-b}).

In order to verify the applicability of the sum-of-potentials approximation
we have carried out at a consistent level (cc-pV5Z/MP4) a series of
ab-initio calculation on the Li$_{2}^{+}$He$_{2}$ system for which
such 3-body effects may arise. The analysis carried out here is analogous
to what we have already reported in Ref. \cite{zb:bodo05-b} for the
LiH$^{+}$ dopant. The 3-body forces should mainly originate from
the induced multipoles (attractive and repulsive contributions) and
by the weaker Axilrod-Teller effects \cite{cl:lenzer01}. For our
preliminary study, we have therefore chosen four different geometries
of the complex where the two helium atoms are close enough to the
ion molecule in order to contribute significantly to 3-body forces.
The first geometry (geometry A) is weakly repulsive ($\sim70$ cm$^{-1}$)
with the two helium atoms at 3.11~\AA~and is the upper one reported
in the inset of Table 3. The second (B) and the third (C) geometries
are similar in shape to the second one sketched in the same table,
but they differ for the values of the distances between the atoms.
The last one (geometry opt) is also similar in shape to the latter,
but it comes instead from a full minimization at the MP2/cc-pv5Z level
of the entire complex: all the relevant distances are reported in
Table 3. %
\begin{table}

\caption{Distances (\AA) and energies (cm$^{-1}$) of the test geometries
used to estimate the importance of 3-body effects in the ionic system.
Energies have been corrected for BSSE. }

\begin{center}\includegraphics[scale=0.5]{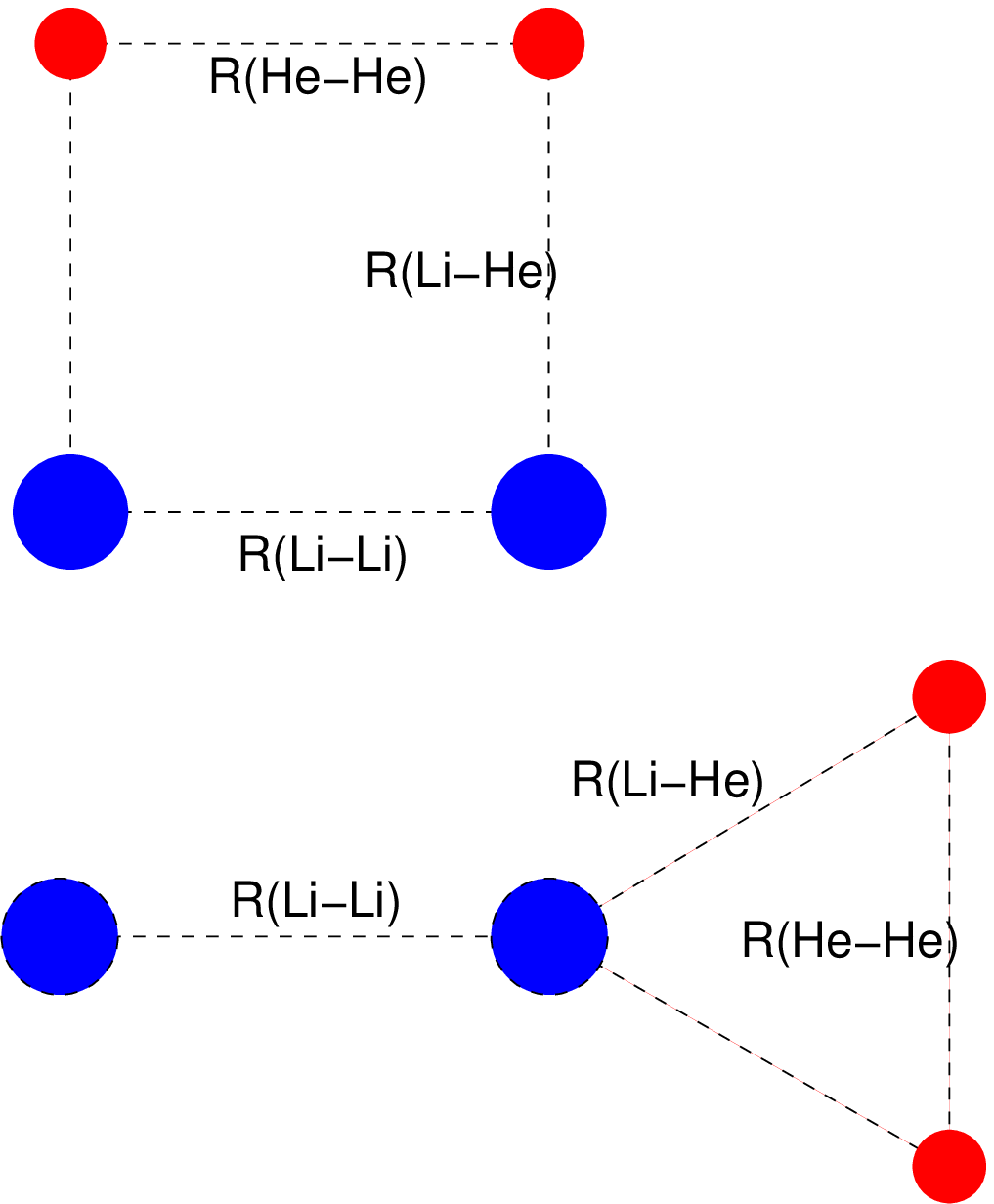}\end{center}

\begin{center}\begin{tabular}{|c|c|c|c|c|}
\hline 
Quantity&
Geom A&
Geom B&
Geom C&
Geom opt\\
\hline
\hline 
R(Li-Li)&
3.11&
3.11&
3.11&
3.04382\\
\hline 
R(Li-He)&
3.00&
3.00&
3.60&
1.91530\\
\hline 
R(He-He)&
3.11&
3.00&
1.25&
2.52866\\
\hline 
$V${[}Li$_{2}^{+}$He$_{2}${]}&
66.113&
-223.547&
11376.770&
-811.745\\
\hline 
$V${[}Li$_{2}^{+}$He{]}&
38.745&
-110.088&
-52.863&
-404.729\\
\hline 
$V${[}He$_{2}${]}&
-3.436&
-2.9471&
11367.744&
22.8860\\
\hline 
$V$(3B)&
-7.941 (12\%)&
-0.424 (0.2\%)&
114.752 (1\%)&
-25.173 (3.1\%)\\
\hline
\end{tabular}\end{center}
\end{table}
The analysis of 3-body forces is done in the following way: for each
geometry we calculate the interaction energy of the entire Li$_{2}^{+}$He$_{2}$
complex, of one of its 2-body fragments (there are two identical ones
in symmetrical geometries) Li$_{2}^{+}$He and of the remaining He$_{2}$
dimer by using the following formulae where the geometry is fixed,
it is the same in each fragment and the counterpoise correction is
used for each interaction energy: 
\begin{eqnarray}
V[\mathrm{Li_{2}^{+}}\mathrm{He_{2}}] & = & E[\mathrm{Li_{2}^{+}He_{2}}]
-E[\mathrm{Li_{2}^{+}}]-2E[\mathrm{He}]\nonumber \\
V[\mathrm{Li_{2}^{+}}\mathrm{He}] & = & E[\mathrm{Li_{2}^{+}}\mathrm{He}]
-E[\mathrm{Li_{2}^{+}}]-E[\mathrm{He}]\nonumber \\
V[\mathrm{He_{2}}] & = & E[\mathrm{He_{2}}]-2E[\mathrm{He}]
\end{eqnarray}
the residual 3-body interaction is simply calculated by using the
expression
\begin{equation}
V[\mathrm{Li_{2}^{+}He_{2}}]-2V[\mathrm{Li_{2}^{+}He}]-V[\mathrm{He}_{2}].
\end{equation}
As can be seen from Table 3, the 3-body interaction is always a small
percentage of the total interaction in the various geometries except
for geometry A where it represents more than 10\% of the total interaction.
It is however important to note that the geometry A corresponds to
a repulsive geometry for the Li$_{2}^{+}$-He fragment and therefore
it is not relevant for optimization purposes.

\section{Conclusions}

We have computed two accurate Potential Energy Surfaces for two different
systems that are of interest for the experimental and theoretical
study of helium droplets doped with alkali metal molecules. The two
molecules considered here are the triplet state of Li$_{2}$ and the
ground state ionic Li$_{2}^{+}$. As should be expected, we found
the two PES's to be markedly different:

\begin{itemize}
\item the interaction of the neutral moiety is similar to the one we have
already studied for ground state singlet Li$_{2}$ \cite{zb:bodo04,zb:bodo04-b}:
a very weak interaction that confirms once more the tendency of high
spin compounds of alkali metals to reside on the surface of helium
droplets. A modified version of the same interaction that gives rise
to slightly deeper potentials can be obtained following our earlier
proposal in Ref. \cite{zb:bodo05-a} and also found to yield weaker
interaction potentials than those between He partners.
\item the triplet dimer interaction with helium is showing here a markedly
different anisotropy from that found earlier on for the singlet state
of Li$_{2}$: the minimum energy configuration is, in fact, given
by T-shaped structures as opposed to the linear structures obtained
for the singlet interaction. 
\item the ionic interaction, instead, is much stronger and more orientation-dependent:
it should therefore lead to full solvation of the molecular moiety
inside the droplets.
\item a preliminary analysis of the three-body effects on the interactions
with more He atoms indicates that such effects are relatively small
and should therefore allow the use of an approximate description of
the full potential energy landscapes in $\mathrm{Li}_{2}^{+}\mathrm{He_{n}}$
clusters in terms of two-body potentials.
\end{itemize}
Both surfaces have been fitted using rather simple analytical expression
reported in the present paper and which therefore provide working
quality interactions potentials for the title systems. We believe
that such potentials are an important step in the modelling of the
larger clusters behavior because they may be used to set up the total
interaction for much larger systems whenever using the sum-of-potential
approach is found to be a realistic alternative: our present study
and preliminary analysis do seem to suggest that this may be the case
for both the present system. We shall verify such a possibility via
our ongoing calculations for the doped $^{4}\mathrm{He}$ clusters. 

\begin{acknowledgments}
We acknowledge financial support from Rome {}``La Sapienza'' Scientific
Committee, the CASPUR supercomputing Center, the MUIR National Projects
FIRB and PRIN. We also acknowledge support from the INTAS grant 03-51-6170.
\end{acknowledgments}

\end{document}